\documentclass[conference]{IEEEtran}
\IEEEoverridecommandlockouts
\usepackage{cite}
\usepackage{amsmath,amssymb,amsfonts}
\usepackage[version=4]{mhchem}
\usepackage{enumitem}
\usepackage{algorithmic}
\usepackage{algorithm}
\usepackage{graphicx}
\usepackage{textcomp}
\usepackage{xcolor}
\usepackage{hyperref}
\hypersetup{colorlinks = true, 
  linkcolor = blue, 
  urlcolor = blue,
citecolor = blue} 
\def\BibTeX{{\rm B\kern-.05em{\sc i\kern-.025em b}\kern-.08em
T\kern-.1667em\lower.7ex\hbox{E}\kern-.125emX}}
\newcommand{\algref}[1]{Algorithm~\ref{#1}}
\newcommand{\figref}[1]{Fig.~\ref{#1}}
\newcommand{\tabref}[1]{Table~\ref{#1}}
\newcommand{\eqqref}[1]{Eq.~\eqref{#1}}
\DeclareRobustCommand{\IEEEauthorrefmark}[1]{\smash{\textsuperscript{\footnotesize #1}}}
\begin{document}
\bstctlcite{IEEEexample:BSTcontrol}
\title{Optimizing Mirror-Image Peptide Sequence Design for Data
Storage via Peptide Bond Cleavage Prediction}
\author{
	\IEEEauthorblockN{
		Yilong Lu\IEEEauthorrefmark{1}, 
		Si Chen\IEEEauthorrefmark{2}, 
		Songyan Gao\IEEEauthorrefmark{2}, 
        Han Liu\IEEEauthorrefmark{3},
		Xin Dong\IEEEauthorrefmark{2},
        Wenfeng Shen\IEEEauthorrefmark{3}$^{,*}$
        and Guangtai Ding\IEEEauthorrefmark{1}$^{,*}$
        }
        
        \IEEEcompsocitemizethanks{\IEEEcompsocthanksitem $^*$corresponding authors: Guangtai Ding(gtding@shu.edu.cn)
        \IEEEcompsocthanksitem Wenfeng Shen(wfshen@sspu.edu.cn)
        }
	\IEEEauthorblockA{\IEEEauthorrefmark{1}\textit{School of Computer Engineering and Science}, \textit{Shanghai University}, Shanghai, China}
	\IEEEauthorblockA{\IEEEauthorrefmark{2}\textit{School of Medicine}, \textit{Shanghai University}, Shanghai, China}
	\IEEEauthorblockA{\IEEEauthorrefmark{3}\textit{School of Computer and information Engineering}, \textit{Institute for Artificial Intelligence},\\ \textit{Shanghai Polytechnic University}, Shanghai,  China} 
	
    }

\maketitle

\begin{abstract}
Traditional non-biological storage media, such as hard drives, face limitations in both storage density and lifespan due to the rapid growth of data in the big data era. Mirror-image peptides composed of D-amino acids have emerged as a promising biological storage medium due to their high storage density, structural stability, and long lifespan.
The sequencing of mirror-image peptides relies on \textit{de-novo} technology. However, its accuracy is limited by the scarcity of tandem mass spectrometry datasets and the challenges that current algorithms encounter when processing these peptides directly.
This study is the first to propose improving sequencing accuracy indirectly by optimizing the design of mirror-image peptide sequences. In this work, we introduce DBond, a deep neural network based model that integrates sequence features, precursor ion properties, and mass spectrometry environmental factors for the prediction of mirror-image peptide bond cleavage. In this process, sequences with a high peptide bond cleavage ratio, which are easy to sequence, are selected.
The main contributions of this study are as follows. First, we constructed MiPD513, a tandem mass spectrometry dataset containing 513 mirror-image peptides. Second, we developed the peptide bond cleavage labeling algorithm (PBCLA), which generated approximately 12.5 million labeled data based on MiPD513. Third, we proposed a dual prediction strategy that combines multi-label and single-label classification. On an independent test set, the single-label classification strategy outperformed other methods in both single and multiple peptide bond cleavage prediction tasks, offering a strong foundation for sequence optimization.
\end{abstract}


\section{Introduction}
\begin{figure*}[htbp]
  \centerline{\includegraphics[scale=1]{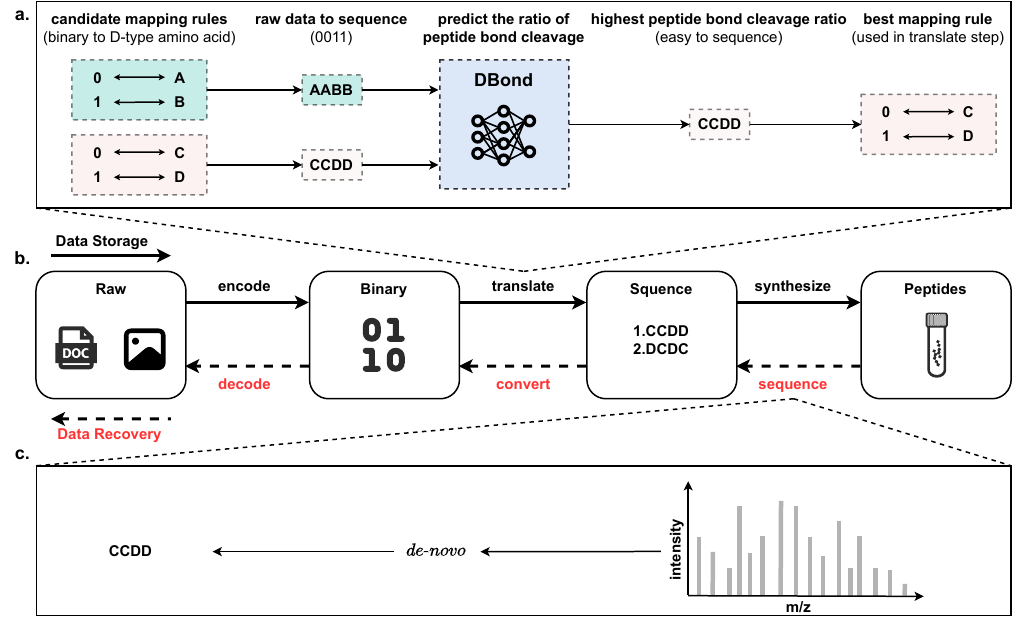}}
  \caption{Overview of data storage technology based on mirror-image peptide.
    \textbf{(a)} The peptide bond cleavage ratio predicted by DBond can be used to identify sequences that are easier to sequence, thereby finding the optimal mapping rules and optimizing sequence design.
    \textbf{(b)} The Data storage technology based on mirror-image
    peptide sequences can be divided into 2 stages: data storage and
    data recovery, further categorized into 6 steps.
    \textbf{(c)} During the sequencing of mirror-image peptides, \textit{de-novo} methods are required to accurately identify the corresponding D-amino acid sequence for each specific mirror-image peptide.}
  \label{fig1}
\end{figure*}
Technological advancements have ushered humanity into the era of big data. In 2010, the total volume of global data was approximately 2 ZB, and it is projected to reach 394 ZB by 2028\cite{ref01}. Nearly all data have been stored in digital formats, since the invention of electronic devices in the last century\cite{ref02}. Currently, magnetic tapes and hard drives are commonly used data storage media, while magnetic tapes primarily used for storing large volumes of infrequently accessed data. However, tape storage density has nearly reached its physical limit. Moreover, tapes require regular replacement since they typically retain data for only 10 to 20 years\cite{ref03}. As a result, with data volumes growing relentlessly, the cost of tape-based storage continues to rise, fueling demand for more affordable storage solutions.

In recent years, a new generation of data storage technologies based on biological macromolecules has been rapidly evolving, offering solutions to many of the limitations of traditional storage devices. For example, when using DNA as a storage medium, the data storage density can reach up to 295 PB/g, and data can be preserved for 20,000 years at 9.4 $^{\circ}$C without any protection\cite{ref04}.
Peptides are biological macromolecules similar to DNA.
Compared with DNA, peptides exhibit more complex biological structures and greater stability. Peptide-based data storage technology offers higher storage density and longer lifespan\cite{ref03}. Peptides can be categorized into two types based on their amino acid composition: natural peptides (composed of L-amino acids) and mirror-image peptides (synthesized from D-amino acids). Mirror-image peptides are particularly ideal for high-density, long-term data storage\cite{ref05}, as their enhanced stability stems from the inability of natural enzymes to degrade them. This inherent resistance ensures reliable preservation of encoded information.

The basic workflow of mirror-image peptide–based data storage technology is illustrated in \figref{fig1}, where one of the key steps is to sequence the mirror-image peptides\cite{ref03, ref05}.
The key objective of sequencing is to accurately determine the D-amino acid sequence of the peptide. Accurate data recovery is not possible if the sequencing performance is poor, as the corresponding D-amino acid sequence of the mirror-image peptide cannot be reliably identified.

\textit{De-Novo} sequencing algorithms have unique advantages in the field of biological data storage due to their ability to sequence peptides without relying on databases\cite{ref03, ref05}.
Early \textit{de-novo} sequencing algorithms primarily relied on exhaustive search strategies\cite{ref07} and graph theory–based approaches\cite{ref08, ref09, ref10, ref11, ref12}. However, as the volume of data has continued to grow, these methods have encountered significant performance bottlenecks.
Machine learning–based methods have been applied into the field to address these challenges. Notable examples include NovoHMM\cite{ref13}, which is based on a Hidden Markov Model, and Novor\cite{ref14}, which employs a decision tree model. Neural network-based methods are also widely used to further improve sequencing performance. Network models including recurrent neural networks (RNNs), convolutional neural networks (CNNs), and transformers have achieved outstanding results.
Among them, Peakonly\cite{ref18} uses CNN to distinguish between real peaks and noise peaks in the tandem mass spectra, thereby improving the accuracy of downstream sequencing workflows. Casanovo\cite{ref21} uses the transformer framework to directly map from mass spectra to amino acid sequences.
Other related works include\cite{ref15, ref16, ref17, ref18, ref19, ref20, ref21} etc. Research by Muth et al. \cite{ref22} and Bealie et al. \cite{ref23} has demonstrated that deep learning–based \textit{de-novo} sequencing methods outperform traditional algorithms across multiple datasets.

Mirror-image peptides used for data storage have several key characteristics: first, they are composed of D-enantiomers of both natural amino acids and unnatural amino acids; second, mirror-image peptide datasets are relatively scarce; third, higher data storage densities correspond to longer mirror-image peptide sequences\cite{ref03, ref05}. Therefore, the performance of \textit{de-novo} sequencing methods based on deep learning is very limited on mirror-image peptide datasets, even though these algorithms have achieved remarkable results on natural peptide datasets. This limitation constrains the accuracy of mirror-image peptide sequencing and, in turn, hinders the advancement of mirror-image peptide–based data storage technologies.
\begin{figure*}[hbtp]
  \centerline{\includegraphics[scale=1]{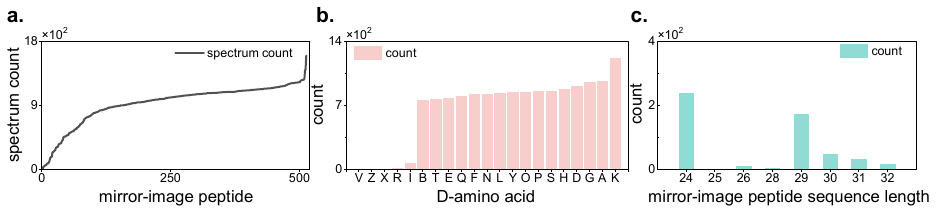}}
  \caption{Statistical information of MiPD513.
    \textbf{(a)} The x-axis represents the types of mirror-image peptides, while the y-axis indicates the number of tandem mass spectra.
    \textbf{(b)}The x-axis represents the sequence lengths of mirror-image peptides, while the y-axis indicates the number of mirror-image peptides.
    \textbf{(c)} The x-axis represents the types of D-amino acids, while the y-axis indicates to the number of mirror-image peptides.}
  \label{fig2}
\end{figure*}
To address this issue, we propose selecting the optimal mapping rule and optimizing the mirror-image peptide sequence design during the translate step in \figref{fig1}(b). This ensures that the resulting sequences are easier to sequence, thereby indirectly enhancing overall sequencing performance. Existing studies have shown that peptides are easier to sequence when each amino acid residue is supported by at least one peak in the tandem mass spectrum\cite{ref24}. The number of cleaved peptide bonds can be used to represent the number of amino acid residues in the tandem mass spectrum. Therefore, we propose using the ratio of cleaved peptide bonds in the mirror-image peptide as an indicator of the sequencing difficulty for the mirror-image peptide. Multiple mapping rules can exist between raw data and individual D-amino acids, allowing the same raw data to be encoded into different mirror-image peptide sequences, as illustrated in \figref{fig1}(a). In practice, the selection of a specific mapping rule often relies on the experience of researchers\cite{ref03, ref05}.
By predicting the peptide bond cleavage ratios of candidate mirror-image peptide sequences, we can identify the optimal mapping rule between raw data and D-amino acids that minimizes the overall sequencing difficulty. This approach indirectly improves sequencing performance. To achieve the above objectives, this study primarily accomplished the following:

(1). A tandem mass spectrometry dataset of mirror-image peptides (MiPD513) was constructed, which includes 513 types of mirror-image peptides and a total of 477, 669 tandem mass spectra.

(2). An automated peptide bond cleavage labeling algorithm (PBCLA) was developed to automatically extract peptide bond cleavage information from tandem mass spectra. Using this method, a total of 12, 473, 724 labeled instances were generated from the MiPD513 dataset, covering 303 distinct types of peptide bonds.

(3). A deep neural network based model DBond was proposed, which integrates sequence features, precursor ion properties, and mass spectrometry environmental factors for the prediction of mirror-image peptide bond cleavage.

(4). To predict peptide bond cleavage, we explored two strategies. The first treats the problem as a multi-label classification task, predicting the cleavage status of all peptide bonds in a single mirror-image peptide simultaneously. The second decomposes the task into multiple single-label classification tasks, predicting the cleavage status of each peptide bond sequentially. Experimental results show that the second strategy outperforms the first.

\section{Materials and methods}

\subsection{Preliminaries}
In this work, we proposes to indirectly improve sequencing performance by optimizing the sequence design of mirror-image peptides. The optimization process can be described by the following formula:
\begin{equation}
h^\ast = \mathop{arg\space max}\limits_{h} \sum_{d\in \mathcal{D}}g(h(d)) \label{eq1}
\end{equation}
Where $\mathcal{D}$ represents the set of raw data that needs to be mapped to mirror-image peptide sequences, and $\mathcal{H}$ represents the set of mapping rules between raw data and D-type amino acids.In practice, the construction of $\mathcal{H}$ is typically guided by domain knowledge or prior experience.
A specific mapping rule 
$h \in \mathcal{H}$ can be used to map an element $d \in \mathcal{D}$ to a corresponding mirror-image peptide sequence. The function $g$ is used to evaluate the sequencing difficulty of a given mirror-image peptide. A higher value of $g$ indicates that the sequence is easier to sequence.The goal of sequence design optimization is to find an optimal mapping rule $h^\ast$ so that the data in D can be most easily sequenced after being encoded into a mirror-image peptide sequence. 

The peptide bond cleavage ratio can serve as an indicator for evaluating the sequencing difficulty of mirror-image peptides. For a mirror-image peptide sequence $seq$ with a length of $l$, $g$ can be defined as:
\begin{equation}
g(seq) = \frac{1}{l-1}\sum_i^{l-1}y_i\label{eq2}
\end{equation}
Where $\boldsymbol{y}=\{y_i\mid y_i\in\{0, 1\}, 1\le i\le l-1\}$ represents the cleavage status of each peptide bond in the corresponding sequence, and $y_i$ takes 1 when the peptide bond is cleaved, otherwise it takes 0. In practice, determining $\boldsymbol{y}$ requires analyzing tandem mass spectrometry results, which can be expensive and time-consuming. This study proposes to predict $\boldsymbol{y}$ using a deep learning model, with the predicted values denoted as $\boldsymbol{\hat{y}}$.
Considering that the value of $g(seq)$ is discrete, the cross entropy function is used to measure the difference between the predicted $\boldsymbol{\hat{y}}$ and $\boldsymbol{y}$, as shown below:
\begin{equation}
\mathcal{L}(\boldsymbol{y},\boldsymbol{\hat{y}})= \frac{1}{l-1}\sum_i^{l-1}(y_i\log(\hat{y_i})+(1-y_i)\log(1-\hat{y_i}))
\label{eq3}
\end{equation}
Then the optimization goal of deep learning can be expressed as:
\begin{equation}
\theta^\ast = \mathop{arg\space min}\limits_{\theta}\space \mathcal{L}(\boldsymbol{y},\boldsymbol{\hat{y}})
\label{eq4}
\end{equation}
where $\theta^{\ast}$ represents the optimal parameters for the model.
\subsection{Mirror-Image peptide dataset}

The mirror-image peptide dataset MiPD513 contains 513 mirror-image peptides, each composed entirely of D-amino acids, and was synthesized by 
the School of Medicine at Shanghai University.
In addition to the 20 common amino acids, several special amino acids such as D-Dap($\ce{C3H8N2O2}$, B), D-Orn($\ce{C5H12N2O2}$, O), 3-(3-Pyridyl)-D-Ala($\ce{C8H10N2O2}$, X), and D-Cha($\ce{C9H17NO2}$, Z) were incorporated during synthesis. For tandem mass spectrometry analysis, each peptide sample was prepared at a concentration of 10$\mu g/ml$ and analyzed using the Thermo Fisher Vanquish UPLC and QEXACTIVE PLUS Mass Spectrometer. High-energy collisional dissociation (HCD) was employed as the fragmentation method, and multiple experiments were conducted under varying normalized collision energies (NCE). This process generated a total of 477, 669 tandem mass spectra, which were subsequently processed using MSConvert\cite{ref25}. Additional relevant information about the dataset is illustrated in \figref{fig2}.

\subsection{Peptide bond cleavage labelling algorithm}
We proposes the Peptide Bond Cleavage Labeling Algorithm (PBCLA) to extract cleavage information from raw tandem mass spectra. PBCLA involves two main steps. The first step matches fragment ions based on their mass-to-charge ratios ($m/z$) and intensities in the tandem mass spectrum.
Only 6 types of fragment ions are considered: $\mathrm{b}, \mathrm{y}, \mathrm{b}\text{–}\ce{H2O}, \mathrm{b}\text{–}\ce{NH3}, \mathrm{y}\text{–}\ce{H2O}, \mathrm{y}\text{–}\ce{NH3}$. The matching process allows a maximum fragment ion charge state of 2 and uses an $m/z$ tolerance of 20 ppm.

The second step involves calculating the cleavage status of each peptide bond based on the fragment ion information obtained from the first step. According to the calculation results, if the peptide bond is cleaved, it is marked as a positive sample, otherwise it is marked as a negative sample. Let $S=\{(mz_i, intensity_i)\mid 1\leq i\leq n\}$ denote the raw tandem mass spectrum corresponding to a mirror-image peptide sequence $seq$ of length $l$, where $mz_i$ and $intensity_i$ represent the $m/z$ and absolute intensity of the$i$-th data point in the raw spectrum, respectively.
Define the possible charge of the fragment ion as $C=\{1, 2\}$and the possible type of the fragment ion as $T=\{\mathrm{b}, \mathrm{y}, \mathrm{b}\text{–}\ce{H2O}, \mathrm{b}\text{–}\ce{NH3}, \mathrm{y}\text{–}\ce{H2O}, \mathrm{y}\text{–}\ce{NH3}\}$. Then, the set of all theoretically possible fragment ions generated by$seq$can be defined as: $I^t=\{(mz_j, charge_j, residue_j, type_j)\mid 1\leq j\leq m, charge_j\in C, 1\leq residue_j\leq l-1, type_j\in T\}$.
The set of fragment ions matched from the raw mass spectrum is denoted as$I^e$, where $I^e\subseteq I^t$. The cleavage labels of each peptide bond are defined as: $Y=\{y_k\mid 1\leq k\leq l-1, \; y_k\in\{0, 1\}\}$.
Based on these definitions, the pseudocode for the fragment ion matching algorithm applied to raw tandem mass spectra is presented in \algref{alg1}, and the pseudocode for PBCLA is shown in \algref{alg2}.

\begin{algorithm}[!h]

  \caption{Fragment ion matching algorithm}
  \label{alg1}
  \textbf{Input:}
  mirror-image peptide sequence $seq$,
  sequence length $l$,
  tandem mass spectrum $S$,
  fragment ion charge $C$,
  fragment ion type $T$,
  matching error $ppm$,
  function used to calculate the theoretical $m/z$ of fragment ions $f$
  \par\textbf{Output:} matched fragment ion $I^e$
  \par
  \begin{algorithmic}[1]

    \STATE $j\leftarrow 1$
    \FORALL{$charge \in C$ }
    \FORALL{$type \in T$ }
    \FOR{$residue=1$  \textbf{to}  $l-1$ }
    \STATE $mz\leftarrow f(charge,type,residue,seq)$
    \STATE $I^t[j] \leftarrow (mz,charge,residue,type)$
    \STATE $j\leftarrow j+1$
    \ENDFOR
    \ENDFOR
    \ENDFOR
    \STATE $j\leftarrow 1$
    \FORALL{$ion \in I^t$ }
    \STATE find the $mz_i$ from $S$ that is closest to $ion.mz$
    within the $ppm$ error range
    \IF{$1\le i \le n$}
    \STATE $I^e[j] \leftarrow ion$
    \STATE $j\leftarrow j+1$
    \ENDIF
    \ENDFOR
    \STATE \textbf{output} $I^e$
  \end{algorithmic}
\end{algorithm}

\begin{algorithm}[!h]

  \caption{Peptide bond labeling algorithm}
  \label{alg2}
  \textbf{Input:}
  mirror-image peptide sequence $seq$,
  sequence length $l$,
  \algref{alg1} output $I^e$
  \par\textbf{Output:} peptide bond label $Y$
  \par
  \begin{algorithmic}[1]
    \STATE $T^b \leftarrow
    \{\mathrm{b},\mathrm{b\text{-}\ce{H2O}},\mathrm{b\text{-}\ce{NH3}}\}$
    \STATE $T^y \leftarrow
    \{\mathrm{y},\mathrm{y\text{-}\ce{H2O}},\mathrm{y\text{-}\ce{NH3}}\}$
    \FOR{$residue=1$  \textbf{to}  $l-1$ }
    \STATE $index_b \leftarrow residue$
    \STATE $index_y \leftarrow l-1-residue$
    \STATE $Y[residue]\leftarrow 0$
    \FORALL{$ion \in I^e$ }
    \IF{$ion.residue = index_b$  \AND $ion.type \in T^b $}
    \STATE $Y[residue] \leftarrow 1$
    \ENDIF
    \IF{$ion.residue = index_y$  \AND $ion.type \in T^y $}
    \STATE $Y[residue] \leftarrow 1$
    \ENDIF
    \ENDFOR
    \ENDFOR
    \STATE \textbf{output} $Y$
  \end{algorithmic}
\end{algorithm}

\subsection{Grouping of features}
During model training, features with different semantic information are fed into the neural network. These features are grouped based on prior knowledge to support more effective learning, allowing the network to apply suitable modules tailored to the characteristics of each feature group.

The first set of features, referred to as $state$ features, includes the precursor ion charge, precursor ion $m/z$, and the absolute intensity of the precursor ion. In mass spectrometer, peptides are first ionized, acquiring a specific charge and exhibiting properties such as intensity. Ionization can alter interactions between amino acids within the peptide due to differences in charge states, leading to precursor ions with distinct characteristics\cite{ref26}. State features are used to represent peptides under these specific conditions.
The second set of features is referred to as $bond$ features, which include the relative position of the peptide bond in the sequence, counted from the N-terminus. The intensity of ion fragments is influenced by the corresponding residue\cite{ref27}, which in turn is affected by the position of the cleaved peptide bond. The third set of features is referred to as $env$ features, which include collision energy and the mass spectrometry scan number. Tandem mass spectrometry is performed under specific collision energies and involves multiple consecutive scans. These features describe the experimental environment.
The fourth set of features is the $sequence$ feature, which refers to the mirror-image peptide sequence itself. This feature accounts for the influence of sequence composition on peptide bond cleavage.

\subsection{The architecture of the DBond model}
\begin{figure}[htbp]
  \centerline{\includegraphics[scale=1]{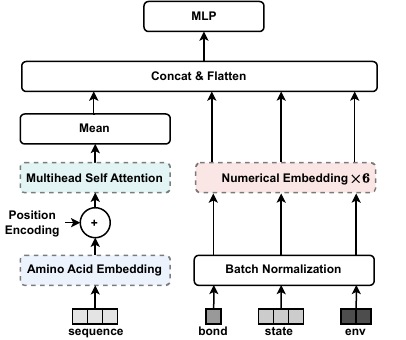}}
  \caption{The overall architecture of DBond. By
    adjusting the output dimensions of the MLP layer, it can be applied
    to both single-label classification tasks and multi-label
  classification tasks.}
  \label{fig3}
\end{figure}

Based on deep learning methods, we developed the DBond model, whose overall architecture is illustrated in \figref{fig3}. The mirror-image peptide sequence $seq$ is composed of D-amino acids represented by single-letter codes. The types and relative positional relationships of these D-amino acids determine the physicochemical properties of the mirror-image peptide, which in turn influence peptide bond cleavage.

The multi-head self-attention mechanism (MSA) is employed to learn dependencies among D-amino acids and to extract information from the mirror-image peptide sequence. Given a mirror-image peptide sequence of length $l$, $\boldsymbol{seq}=(aa_1, aa_2, \ldots, aa_l)$, where $aa_i\in\mathcal{A}$ denotes the $i$-th D-amino acid and $\mathcal{A}$ is the alphabet of D-amino acids, the feature construction process of the mirror-image peptide can be formally expressed as follows:
\begin{equation}
E_{seq} = MSA(embed(\boldsymbol{seq})+pe(\boldsymbol{seq})) \label{eq5}
\end{equation}
Here, $E_{seq}\in\mathbb{R}^{l\times d}$ represents the feature embeddings of the mirror-image peptide sequence, where$d$is the embedding dimension for each D-amino acid. $MSA(\cdot)$ denotes the multi-head self-attention encoder, $embed(\cdot)$ represents the embedding function for amino acids, and $pe(\cdot)$ is the positional encoding function.
Since the $state$, $bond$, and $env$ features influence peptide bond cleavage in different ways, DBond embeds these features separately to capture their distinct effects. Let $\boldsymbol{x}=(x_1, x_2, \ldots, x_n)$ represent the numerical input features such as $state$, $bond$, or $env$, where $x_i\in\mathbb{R}$ and $n$ is the length of the feature vector. The embedding process for the numerical features $\boldsymbol{x}$ can be expressed as:
\begin{equation}
E_x = ReLU(L(bn(\boldsymbol{x}))) \label{eq6}
\end{equation}
Here, $E_x\in\mathbb{R}^{n\times d}$ represents the high-dimensional embedding of $\boldsymbol{x}$ after the embedding process. $L(\cdot)$ denotes an affine transformation function, and $bn(\cdot)$ represents batch normalization.
After embedding the input features, the output of DBond, denoted as $\boldsymbol{y}\in\mathbb{R}^m$, can be expressed as:
\begin{equation}
\boldsymbol{y}=\phi(MLP(\delta(mean(E_{seq}),E_{bond},E_{state},E_{env}))) \label{eq7}
\end{equation}
Here, $m$ denotes the output dimension, $\phi(\cdot)$ represents the sigmoid function, and $MLP$ refers to a multilayer perceptron. The function $\delta(\cdot)$ concatenates the input data along the feature dimension and then flattens it into a vector, while $mean(\cdot)$ computes the mean of the input data along the feature dimension.
\subsection{Experimental setup}

\paragraph{Dataset preprocessing and splitting} After applying PBCLA to MiPD513, the dataset is split into training and test sets at a ratio of 8: 2. Peptide sequences in the test set are excluded from the train set. A 5-fold cross-validation strategy is also employed.
\paragraph{Prediction strategies} Two prediction strategies are proposed in this work to predict the cleavage of each peptide bond in mirror-image peptides. The first strategy formulates the task as a multi-label classification problem, directly predicting the cleavage status of all peptide bonds simultaneously. The second strategy treats it as a set of independent single-label classification problems, sequentially predicting the cleavage status of each peptide bond to determine the overall cleavage pattern of the peptide.
\paragraph{Loss and Evaluation Metrics} \eqqref{eq3} is used as the loss function.
Multi-label classification metrics, as defined in\cite{ref32}, are used to evaluate the prediction of all peptide bond cleavages in a mirror-image peptide. These include \textit{example-based} metrics such as subset accuracy, and \textit{label-based }metrics such as precision and recall. Single-label classification metrics, as defined in\cite{ref33}, are used to assess the prediction of individual peptide bond cleavage. These include metrics such as AUC, accuracy, and F1 score.
\paragraph{Baselines} In this work, we did not use traditional machine learning methods (e. g., XGBoost) as baselines, as they are not well-suited to handling the sequence features of mirror-image peptides for the following reasons.
First, the sequences of mirror-image peptides are complex, with large numbers, variable lengths, and high internal dependencies. These characteristics limit the effectiveness of standard categorical encoding methods like one-hot encoding. Second, existing peptide feature extraction tools and models, such as iFeature\cite{ref28}and AlphaFold3\cite{ref29}, do not support the direct processing of mirror-image peptides composed of D-amino acids or non-standard amino acids. Although deep learning models do not require manual feature extraction, research addressing the specific problem in this study is limited, and suitable baseline models are lacking.
To evaluate the performance of the proposed model, DBond is compared with two representative deep learning models: Prosit\cite{ref30}and PredFull\cite{ref31}. Both are designed to predict peptide tandem mass spectra and can be retrained and tested on the MiPD513 dataset. Prosit predicts the intensities of backbone ions (b, y ions) in tandem mass spectra, while PredFull predicts the intensities across all possible $m/z$ values. Although neither model directly predicts peptide bond cleavage, both can do so indirectly by applying PBCLA to their predicted theoretical spectra.

\section{Results and Discussion}

\subsection{Result of peptide bond cleavage labelling algorithm}
\begin{figure*}[htbp]
  \centerline{\includegraphics[scale=1]{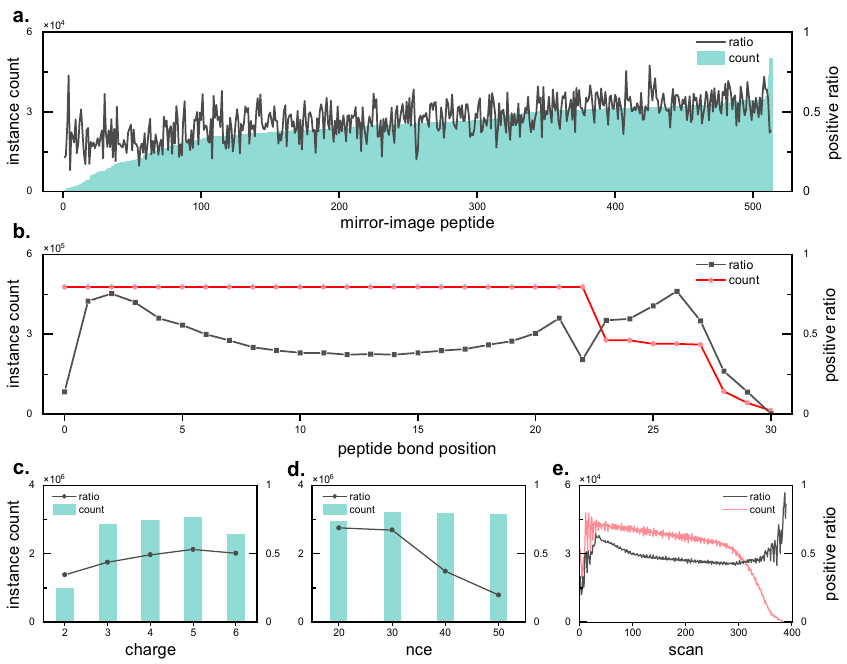}}
  \caption{Labelling results of the PBCLA on MiPD513.
    \textbf{(a)} The x-axis represents the types of mirror-image peptides, the left y-axis indicates the corresponding sample count, and the right y-axis shows the corresponding positive sample ratio (the same applies below).
    \textbf{(b)} The x-axis represents the position of the peptide bond.
    \textbf{(c)} The x-axis represents the charge state of the precursor.
    \textbf{(d)} The x-axis represents the normalized collision energy.
    \textbf{(e)} The x-axis represents the scan number during the tandem mass spectrometry process.}
  \label{fig4}
\end{figure*}
A total of 12,473,724 labeled data were generated by applying PBCLA to the raw tandem mass spectra. Among these instances, those labeled as cleaved peptide bonds, referred to as positive instances, account for approximately 48.03\%. The distribution of sample counts across different peptide sequences, along with their corresponding positive ratios, is shown in \figref{fig4}(a).
Peptide bond positions are indexed starting from the N-terminus, with the first bond labeled as 0, the second as 1, and so on. Instances can be grouped based on these bond positions. The corresponding sample counts and positive ratios are shown in \figref{fig4}(b), which aligns well with experimental observations\cite{ref03, ref27}.
Other factors related to peptide bond cleavage, such as precursor ion charge, NCE, and scan number, have their corresponding sample counts and positive ratios shown in \figref{fig4}(c), \figref{fig4}(d), and \figref{fig4}(e), respectively.

Peptide bond cleavage in mirror-image peptides during tandem mass spectrometry is influenced by multiple factors, including peptide properties, bond-specific characteristics, and experimental conditions. The labeling algorithm proposed in this study enables automated identification of cleavage events and provides insights into how these factors may affect peptide bond cleavage.

\subsection{Performance on single peptide bond cleavage prediction}
The single-label classification strategy transforms the task of predicting peptide bond cleavage in a mirror-image peptide as a series of independent single-label classification problems. Under this strategy, the model’s performance on individual bond-level predictions directly affects the overall prediction accuracy.
Therefore, this study first evaluates the performance of the DBond model on the single peptide bond cleavage prediction task. DBond can predict the cleavage of a single peptide bond by simply adjusting the output dimension. This model is denoted as DBond-s. \tabref{tab1} reports the performance comparison for predicting single peptide bond cleavage within the dataset.
\begin{table}[htpb]
  \centering
  \caption{performance on single peptide bond cleavage prediction(\%)}
  \label{tab1}
  \resizebox{\columnwidth}{!}{%
    \begin{tabular}{ccccccc}
      \hline
      \textbf{Model} & \textbf{AUC} & \textbf{AP} & \textbf{Acc} &
      \textbf{Pre}$^{\ast}$ & \textbf{Rec}$^{\ast}$ & \textbf{F1}$^{\ast}$ \\ \hline
      predfull &  \textbf{×}$^{\star}$ &
      \textbf{×}$^{\star}$ & 51.92 & 51.56 & 51.47 & 51.01 \\
      prosit & \textbf{×}$^{\star}$ & \textbf{×}$^{\star}$
      & 51.77 & 69.30 & 58.35 & 47.08 \\
      DBond-s & 90.46 & 89.01 &\textbf{82.42} &
      \textbf{82.44} & \textbf{82.47} & \textbf{82.41} \\
      \hline
      \multicolumn{7}{p{0.9\columnwidth}}{$^{\star}$Since the
        outputs of the predfull and prosit are tandem mass spectra
        corresponding to mirror-image peptides rather than
        probabilities of peptide bond cleavage, the AUC and AP metrics
      were not calculated.}\\
      \multicolumn{7}{p{0.9\columnwidth}}{$^{\ast}$These metrics are calculated using the macro-average approach.}
    \end{tabular}%
  }
\end{table}



As shown in \tabref{tab1}, DBond-s achieved an accuracy of 82.42\% and an F1-score of 82.41\% on the test set, significantly outperforming Prosit and Predfull. This indicates that DBond-s exhibits superior performance in predicting the cleavage status of individual peptide bonds on the dataset.
During the experiments, Prosit and PredFull showed poor performance in predicting theoretical mass spectra on the MiPD513 dataset. Specifically, the average spectral angle between Prosit's predicted spectra and the real spectra was 32.33\%, while PredFull achieved an average cosine similarity of only 29.69\%. This underperformance may be attributed to the small size of the MiPD513 dataset and the relatively long peptide sequences, which likely hindered effective model training. As a result, the predicted spectra differed significantly from the actual tandem mass spectra. Consequently, when PBCLA was applied to these theoretical spectra, the resulting cleavage predictions were also poor.

\subsection{Performance on multiple peptide bond cleavage prediction}
For any mirror-image peptide, DBond-s can be used to predict the cleavage of each peptide bond in turn, and finally the cleavage of all peptide bonds can be obtained. DBond can also directly predict the cleavage status of multiple peptide bonds simultaneously by adjusting the output dimension. This variant is referred to as DBond-m. \tabref{tab2} and \tabref{tab3} present the experimental results for predicting multiple peptide bond cleavages.
\begin{table}[htpb]
  \centering
  \caption{Performance on multiple peptide bond cleavage prediction(\%)}
  \label{tab2}
  \resizebox{\columnwidth}{!}{%
    \begin{tabular}{ccccccc}
      \hline
      \textbf{Model} & $\mathbf{Acc}_{subset}$ &
      $\mathbf{Acc}_{example}$ & $\mathbf{Pre}_{example}^{\ast}$ &
      $\mathbf{Rec}_{example}^{\ast}$ & $\mathbf{F1}_{example}^{\ast}$ \\ \hline
      predfull & 0.01 & 25.57 & 49.49 & 38.15 & 43.08 \\
      prosit   & 0.99 & 42.81 & 43.25 & \textbf{91.99} & 58.83 \\
      DBond-m   & 5.50 & 57.02 & 72.42 & 69.35 & 70.84 \\
      DBond-s   & \textbf{6.21} & \textbf{60.02} & \textbf{73.10} &
      73.13 & \textbf{73.10} \\ \hline
       \multicolumn{6}{p{0.9\columnwidth}}{$^{\ast}$These metrics are calculated using the macro-average approach.}
    \end{tabular}%
  }
\end{table}
\begin{table}[htpb]
  \centering
  \caption{Performance on multiple peptide bond cleavage prediction(\%)}
  \label{tab3}
  \resizebox{\columnwidth}{!}{%
    \begin{tabular}{ccccccc}
      \hline
      \textbf{Model}  & $\mathbf{Acc}_{label}$ &
      $\mathbf{Pre}_{label}^{\ast}$ & $\mathbf{Rec}_{label}^{\ast}$ &
      $\mathbf{F1}_{label}^{\ast}$ \\ \hline
      predfull & 64.11 & 43.83 & 33.15 & 36.13 \\
      prosit   & 64.00 & 39.23 & 86.45 & 52.72 \\
      DBond-m   & 85.95 & 66.09 & 65.11 & 65.24 \\
      DBond-s   & \textbf{86.88} & \textbf{67.77} & \textbf{69.71} &
      \textbf{68.34} \\ \hline
      \multicolumn{5}{p{0.9\columnwidth}}{$^{\ast}$These metrics are calculated using the macro-average approach.}
    \end{tabular}%
  }
\end{table}


The experimental results demonstrate that Predfull and Prosit still perform poorly when predicting the cleavage of multiple peptide bonds in mirror-image peptides. However, Prosit achieved the highest recall score, while simultaneously obtaining the lowest precision score. This indicates that in the theoretical spectra predicted by Prosit, the vast majority of ion fragments have intensities greater than 0, which leads to most instances being labeled as positive after applying PBCLA.
A comparison between the experimental results of DBond-s and DBond-m shows that converting the multi-peptide bond cleavage prediction task into multiple single-label classification problems improves performance, despite ignoring label dependencies. This improvement may be due to the limitations of the multi-label formulation: the dataset contains relatively few instances, each associated with many labels, leading to a sparse solution space and reduced learning effectiveness for DBond-m.
The subset accuracy metric measures the proportion of predictions that exactly match the true cleavage pattern across all peptide bonds in a mirror-image peptide.
According to this metric, accurately predicting the cleavage of all peptide bonds in a mirror-image peptide is highly challenging.

\section{Conclusion}
This study proposes using the peptide bond cleavage ratio in mirror-image peptides during tandem mass spectrometry as an indicator of sequencing difficulty.
Sequences with higher cleavage ratios, which suggest easier sequencing, can be selected based on the predicted cleavage status of each peptide bond.
Based on this, optimal mapping rules between raw data and D-amino acids can be identified to guide the design of mirror-image peptide sequences and indirectly improve sequencing performance.

To achieve these objectives, we constructed a tandem mass spectrometry dataset of mirror-image peptides named MiPD513 and proposed a peptide bond cleavage labeling algorithm called PBCLA.
To predict the cleavage status of each peptide bond in a mirror-image peptide, we introduce a deep learning model called DBond, which takes sequence features, precursor state features, and mass spectrometry environmental factors as input.
For the cleavage prediction task, two strategies were employed. One uses multi-label classification, and the other treats the problem as a series of independent single-label classification tasks.
Experimental results show that DBond achieves high predictive performance. The single-label classification strategy performs better and provides valuable guidance for optimizing mirror-image peptide sequences.


\section*{Acknowledgment}
This work was supported by the AI-Driven Reform of Scientific Research Paradigms and Discipline Leapfrogging Initiative (A30YD250115-04). The authors also acknowledge data support from the School of Medicine, Shanghai University, and computational support from SSPU AI Lab.

\bibliographystyle{IEEEtran}
\bibliography{ref.bib}
\vspace{12pt}
\end{document}